\documentclass[%
reprint,
showkeys,
nofootinbib,
amsmath,amssymb,
aps,
prl,
]{revtex4-1}
\usepackage[marginal]{footmisc}
\usepackage{graphicx}
\usepackage{dcolumn}
\usepackage{bm}
\usepackage[colorlinks,allcolors=black,CJKbookmarks=True]{hyperref}

\begin{document}
	
	\title{A New Picture of Cosmic String Evolution and Anisotropic Stochastic Gravitational-Wave Background}
	
	\author{Rong-Gen Cai$^{3,2,1}$}
	\email{cairg@itp.ac.cn}
	
	\author{Zong-Kuan Guo$^{3,2,1}$}
	\email{guozk@itp.ac.cn}
	
	\author{Jing Liu$^{1,2}$}
	\email{corresponding author: liujing@ucas.ac.cn}
	
	\affiliation{$^{1}$School of Fundamental Physics and Mathematical Sciences, Hangzhou Institute for Advanced Study, University of Chinese Academy of Sciences, Hangzhou 310024, China
	}
	
	\affiliation{$^{2}$School of Physical Sciences, University of Chinese Academy of Sciences,
		No.19A Yuquan Road, Beijing 100049, China}
	
	\affiliation{$^{3}$CAS Key Laboratory of Theoretical Physics, Institute of Theoretical Physics,
		Chinese Academy of Sciences, P.O. Box 2735, Beijing 100190, China}
	
	\begin{abstract}
		We investigate  the anisotropies of the stochastic gravitational-wave background (SGWB) produced by cosmic strings associated with the spontaneous U(1) symmetry breaking of Grand Unified Theory, which happens at the onset of inflation. The string network evolution is determined by primordial fluctuations and never reaches the scaling regime. The string loops are inhomogeneously distributed in large scale regions, resulting in large anisotropies in the SGWB. We find that the angular power spectrum of SGWB anisotropies  depends on frequency, which is testable in multiband observations of GWs. In particular, GWs from the cosmic strings can appropriately interpret the common-spectrum process reported by NANOGrav collaboration, and the angular power spectrum in the nanohertz band, $\mathtt{l}(\mathtt{l}+1)C_{\mathtt{l}}=5.6\times 10^{-2}$ at large scales, is expected to be detectable by pulsar timing array experiments in the near future.
		
	\end{abstract}
	
	\maketitle

	\emph{Introduction}. 
	The exploration of stochastic gravitational-wave backgrounds~(SGWBs) from the early Universe can shed lights on new physics at high energy scales and the invisible sector weakly coupled to the standard model of particle physics~\cite{Cai:2017cbj,Bian:2021ini}.
	Recently, the common-spectrum process reported by the NANOGrav collaboration in their 12.5-yr result~\cite{Arzoumanian:2020vkk} receives much attention and is interpreted as GWs sourced in the early Universe~\cite{DeLuca:2020agl,Vaskonen:2020lbd,Bian:2020urb,Cai:2020ovp,Liu:2020mru,Kitajima:2020rpm,Inomata:2020xad}\footnote{The confirmation of the GW detection requires more observational data in the future~\cite{NANOGrav:2020spf}.}.
	The scale-invariant energy spectrum of GWs from cosmic strings could naturally explain the common-spectrum process~\cite{Ellis:2020ena}. The string tension implied by the PTA data coincides well with the prediction of the spontaneous breaking of $U(1)$ symmetry at the Grand Unified Theory~(GUT) scale~\cite{King:2020hyd,King:2021gmj}. In this $Letter$, we find  that the frequency-dependent anisotropies present in the SGWB from cosmic strings, and it is detectable by multiband observations of GWs.
	
	
	Cosmic strings are long-living one-dimensional topological defects predicted as a general consequence of a $U(1)$ symmetry spontaneous breaking~\cite{Nielsen:1973cs,Kibble:1976sj,Vilenkin:1981kz}, and the oscillation of cosmic strings generates detectable GWs ~\cite{Vilenkin:1981bx,Burden:1985md,Scherrer:1990pj,Quashnock:1990wv,Allen:1991bk,Figueroa:2012kw,Jenkins:2018nty,Cui:2019kkd,Sousa:2020sxs}. The previous works mainly focus on the scenario where the $U(1)$ symmetry breaking happens after inflation, the string network evolution soon reaches a scaling regime after the string formation, described by the velocity-dependent one-scale~(VOS) model~\cite{Martins:1996jp,Martins:2000cs}. 
	For Nambu-Goto strings, GW production are mainly from the decay of string loops~\cite{Jones-Smith:2007hib,Matsui:2019obe}, which is produced by collisions between long strings~\cite{Kibble:1984hp,Caldwell:1991jj,DePies:2007bm,Sanidas:2012ee,Sousa:2013aaa}. 
	However, for the symmetry breaking scale $\eta\gtrsim 10^{14}$GeV~\cite{Blasi:2020mfx}, the $U(1)$ symmetry is already broken at the beginning of inflation since the Hawking temperature is constrained by CMB observations as $H_{\mathrm{inf}}/2\pi<7.2\times 10^{12}$GeV~\cite{BICEP:2021xfz}. And the reheating process after inflation is generally incomplete before the temperature of the Universe decreases to $\eta$~\cite{Kofman:1994rk,Kofman:1997yn,Micha:2002ey,Figueroa:2017vfa}, so the $U(1)$ symmetry does not  get restored. In this case, cosmic strings can also form after inflation since quantum fluctuations during inflation kick the scalar field across the potential barrier. Unlike the scaling evolution, the string network evolution and the string loop distribution are determined by primordial fluctuations during inflaton,
	and we find the inhomogeneous distribution of string loops results in frequency-dependent large anisotropies in the SGWB with small multiple parameter $\mathtt{l}$. 
	
	
	For convenience, we choose the speed of light  $c=1$ throughout this $Letter$.
	
	\emph{The model and string loop distribution}.
	Consider a simple model with a broken $U(1)$ symmetry, $V(\Phi)=\frac{\lambda}{4}(|\Phi|^{2}-\eta^{2})^{2}$, where $\Phi$ is a complex scalar field and $\eta\gtrsim 10^{14}$GeV. It is natural to consider the inflationary energy scale to be about $10^{16}$GeV, as suggested by many well-motivated models, such as Starobinsky inflation~\cite{Starobinsky:1980te} and natural inflation~\cite{Freese:1990rb}. The temperature of the Universe suddenly decreases from $10^{16}$GeV to the Hawking temperature $H_{\mathrm{inf}}/2\pi \sim 10^{12}$GeV at the onset of inflation, during which the symmetry breaking takes place. In what follows, we apply the approximation $H_{\mathrm{inf}}$is constant during inflation as the standard slow-roll inflation suggests. The evolution of $\Phi$ during inflation can be described by the sum of classical motion and quantum fluctuations. Here the classical motion refers to that $\Phi$ slowly rolls down the potential $V(\Phi)$ towards the vacuum with $\dot{\Phi}\sim\frac{dV/d\Phi}{3H_{\mathrm{inf}}}$ while quantum fluctuations are quantified as $H_{\mathrm{inf}}/2\pi$ per Hubble time. For a general choice of the coupling constant $\lambda\lesssim 0.1$, the quantum fluctuations are at least three times larger than the classical rolling of $\Phi$,
	we then neglect the influence of $V(\Phi)$ as an approximation, so that the distribution of $\Phi$ can be described by a random walk around its the initial value $\Phi_{i}$, i.e., the solution of the Fokker-Planck function with vanishing potential term~\cite{Espinosa:2015qea}
	\begin{equation}\label{eq:FP}
		\mathcal{P}(\tilde{\Phi}, T)=\sqrt{\frac{2 \pi}{H_{\mathrm{inf}}^{3} (T-T_{i})}} \exp \left[-\frac{2 \pi^{2}|\tilde{\Phi}-\Phi_{i}|^{2}}{H_{\mathrm{inf}}^{3} (T-T_{i})} \right]\,,
	\end{equation}
	where $\mathcal{P}(\tilde{\Phi}, T)$ is the probability that the averaged value of $\Phi$ in a Hubble volume takes the value $\tilde{\Phi}$, $T$ is the cosmic time before the end of inflation, $T_{i}$ satisfies $a(T_{i})H_{\mathrm{inf}}=a(t_{0})H_{0}$ where $t_{0}$ and $H_{0}$ are the cosmic time and Hubble parameter at present and $\Phi_{i}$ is the initial value of $\Phi$ at $T_{i}$.
	Before inflation, the $U(1)$ symmetry is not broken and the vacuum expectation value of $\Phi$ is zero, so it is natural to set $\Phi_{i}$ close to zero. As inflation proceeds, quantum fluctuations gradually kick $\Phi$ away from $\Phi_{i}$, and perturbations of $\Phi$ are frozen outside the Hubble horizon. After inflation, the reheating temperature is generally lower than $\eta$ so that the $U(1)$ symmetry does not restore. When the Hubble parameter becomes smaller than the effective mass of $\Phi$ after inflation, in the regions where $\Phi\neq 0$ the field rolls down and settles in the vacuum $|\Phi|=\eta$, and cosmic strings formation happens only in the regions where $|\Phi|\sim 0$.
	
	The following evolution of the string network and loop distribution function are determined by quantum fluctuations during inflation. As we will see, the string network is always more sparse than that in the scaling regime. 
	One can estimate the probability of $|\Phi|\sim 0$ during inflation to obtain the string length per unit volume after the string formation. Eq.~\eqref{eq:FP} gives the probability of $|\tilde{\Phi}|<|\Delta\Phi|$ as
	\begin{equation}
		\Delta\mathcal{P}=\frac{2\pi}{H_{\mathrm{inf}}^{3}(T-T_{i})}\exp\left[-\frac{2\pi^{2}|\Phi_{i}|^{2}}{H_{\mathrm{inf}}^{3}(T-T_{i})}\right]|\Delta\Phi|^{2}\,,
	\end{equation}
	which also means the volume ratio of the regions where $|\tilde{\Phi}|<|\Delta\Phi|$. Since $\Phi$ changes about $H_{\mathrm{inf}}/2\pi$ over a Hubble horizon scale, $H_{\mathrm{inf}}^{-1}$, we approximately apply $d\Phi/dx=H_{\mathrm{inf}}^{2}/2\pi$, then the radius of the cross section of these regions reads $\frac{|\Delta\Phi|}{H_{\mathrm{inf}}^{2}/2\pi}$.
	The averaged length of these regions per unit volume is obtained as $\Delta\mathcal{P}$ devided by the cross sectional area
	\begin{equation}\label{eq:lengthdensity}
		\mathcal{M}(T)=\frac{H_{\mathrm{inf}}}{2\pi^{2}(T-T_{i})}\exp\left[-\frac{\theta_{i}^{2}}{2H_{\mathrm{inf}}(T-T_{i})}\right]\,,
	\end{equation}
	where $\theta_{i}\equiv \frac{|\Phi_{i}|}{H_{\mathrm{inf}}/2\pi}$. Note that Eq.~\eqref{eq:lengthdensity} only contains the contribution of strings with the comoving curvature radius larger than $\frac{1}{a(T)H_{\mathrm{inf}}}$ since Eq.~\eqref{eq:FP} accounts the probability of the averged $\Phi$ over a Hubble volume. 
	
	After inflation, cosmic strings form in these string-like regions. The motion of cosmic strings is frozen where the curvature radius is larger than the Hubble scale, so $\mathcal{M}(T)$
	given in Eq.~\eqref{eq:lengthdensity} is directly related to the length per unit volume~(length density) of those superhorizon strings
	taking into account the expansion of the Universe
	\begin{equation}\label{eq:stringdensity}
		M(t)=\mathcal{M}(T)\frac{H^{2}(t)}{H_{\mathrm{inf}}^{2}}\,,
	\end{equation}	
	where $t$ is an arbitrary cosmic time after the end of inflation which satisfies $a(t)H(t)=a(T)H_{\mathrm{inf}}$.
	Eq.~\eqref{eq:stringdensity} implies the averaged length of strings per Hubble volume is less than $H_{\mathrm{inf}}^{-1}$, so the characteristic length~(defined as $L\equiv\sqrt{\mu/\rho}$, where $\rho$ is the string energy density) is larger than the Hubble scale. 
	Thus, in this case, the string network is very sparse, and the scaling regime is never reached. 
	
	As the Hubble horizon expands, cosmic strings begin to oscillate and produce GWs in the regions where the curvature radius becomes smaller than $H^{-1}(t)$. 
	The length density of strings which begin to oscillate between $t$ and $t-dt$ is
	\begin{equation}\label{eq:dL}
		dM=M(t-dt)\frac{a^{2}(t-dt)}{a^{2}(t)}-M(t)\,,
	\end{equation}
	or equivalently, $dM=\frac{-d}{a^{2}(t)dt}[a^{2}(t)M(t)]$. 
	The factor $\frac{a^{2}(t-dt)}{a^{2}(t)}$ represents the dilution of cosmic strings due to the expansion of the Universe in $dt$, and
	the first term on the r.h.s. of Eq.~\eqref{eq:dL} represents the length density at $t$ of all strings which is frozen at $t-dt$. Substracting the length density of frozen strings at $t$, one can obtain the length density of strings which begin to oscillate in $dt$.
	
	We assume string loops are produced from self-collisions of those oscillating strings~\footnote{A fair amount of loops form initially from primordial perturbations rather than from the collision of long strings.}. Since the characteristic length is larger than $H^{-1}(t)$, we set the initial loop length to be the Hubble scale $l_{s}\sim H^{-1}(t)$, where $H(t)=\beta/t$ and $\beta=1/2$~($2/3$) in the radiation-dominated~(matter-dominated) era. Thus, the number density of the loops produced in $dt$ is $dN(t)=H(t)dM$.
	
	The length of string loops $l(t)$ decreases with
	time due to GW radiation, $l(t)=l_{s}-\Gamma G \mu\left(t-t_{s}\right)$, where we adopt $\Gamma=50$ as simulations imply~\cite{Blanco-Pillado:2017oxo}. 
	The loops with lengths between $l$ and $l+dl$ are produced during $t_{s}$ and $t_{s}+dt_{s}$ where $t_{s}=(l(t)+\Gamma G\mu t)/(\beta^{-1}+\Gamma G\mu)$ and $dt_{s}=(\beta^{-1}+\Gamma G\mu)^{-1}dl$. Finally, the number density of loops with lengths between $l$ and $l+dl$ is 
	
	\begin{equation}\label{eq:loopdf}
		dN(t_{s}(l,t))=\left(\frac{a(t)}{a(t_{s})}\right)^{3}\frac{2\pi H(t)}{\left(\Gamma G\mu+\beta^{-1}\right)}
		\frac{d\left[L(t_{s}) a^{2}(t_{s})\right]}{a^{2}(t_{s})dt_{s}}dl\,.
	\end{equation}
	The loop distribution function $n(l,t)\equiv dN(t_{s}(l,t))/dl$ denotes the number density of loops per unit length and unit volume.
	
\begin{figure}[t]
		\includegraphics[width=3in]{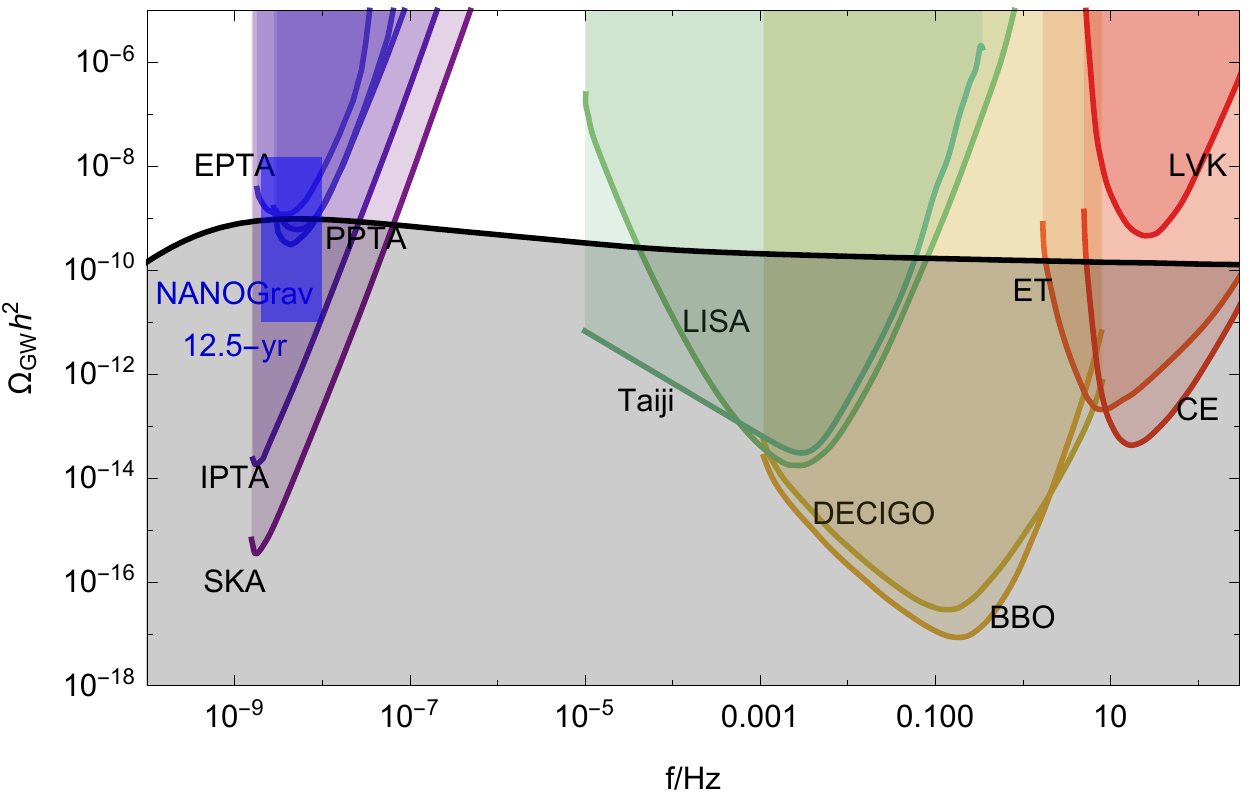}
		\caption{ The black line represents the predicted energy spectrum from cosmic string loops with the loop distribution function Eq.~\eqref{eq:loopdf} and the string tension $G\mu=4\times 10^{-10}$, which naturally explains the common-spectrum process detected by NANOGrav~(the blue box). The sensitivity curves can be found in Ref.~\cite{Schmitz:2020syl}, including EPTA~\cite{Lentati:2015qwp}, PPTA~\cite{Shannon:2015ect}, NANOGrav~\cite{Arzoumanian:2018saf}, IPTA~\cite{Hobbs:2009yy}, SKA~\cite{Carilli:2004nx}, LISA~\cite{Audley:2017drz} Taiji~\cite{Guo:2018npi}, DECIGO~\cite{Kawamura:2011zz}, BBO~\cite{phinney2004big}, LIGO, Virgo and KAGRA~( LVK)~\cite{TheLIGOScientific:2014jea,Somiya:2011np}, CE~\cite{Reitze:2019iox}, ET~\cite{Punturo:2010zz}..
		}
		\label{fig:Omega}
	\end{figure}

	
	\emph{GW production and anisotropies}. 
	The calculation of the GW energy spectrum is based on the loop distribution function Eq.~\eqref{eq:loopdf}.
	We assume cusps dominate the GW production from string loops, then the GW energy spectrum takes the form of the summation of all harmonic modes of string loops~\cite{Allen:1991bk,Binetruy:2009vt}
	\begin{equation}\label{eq:Omega}
		\Omega_{\mathrm{GW}}\left(f\right)=\sum_{j=1}^{\infty} \frac{j^{-4/3}}{\mathcal{E}} \Omega_{\mathrm{GW}}^{j}(f)\,,
	\end{equation}
	where $\mathcal{E}=\sum_{j=1}^{\infty} j^{-4/3}$ is a normalization factor, $j$ denotes the $j$-th harmonic mode, $f=2j/l$ is the GW frequency~\cite{Burden:1985md}, and
	\begin{equation}\label{eq:Omegaj}
		\begin{split}
			\Omega_{\mathrm{GW}}^{j}&(f)=\frac{16 \pi}{3}\left(\frac{G \mu}{H_{0}}\right)^{2} \frac{\Gamma}{f}\times\\ 
			&\int_{t_{f}}^{t_{0}} j n\left(l_{j}\left(t^{\prime}\right), t^{\prime}\right)\Theta(t'-t_{s})\left(\frac{a\left(t^{\prime}\right)}{a(t_{0})}\right)^{5} d t^{\prime}\,,
		\end{split}
	\end{equation}
	where $t_{f}$ is the time of string formation, see Ref.~\cite{Sousa:2020sxs} for details. The Heaviside function $\Theta(t'-t_{s})$ means the string loop starts producing GWs after the loop formation. 
	The predicted GW energy spectrum for $\theta_{i}=0$ and $G\mu=4\times 10^{-10}$ is shown in Fig.~\ref{fig:Omega}, which naturally explains the common-spectrum process detected by NANOGrav.
	The profile of $\Omega_{\mathrm{GW}}(f)$ is almost the same as the previous scaling case. Since the string network is sparse and string loops are less produced, the string tension given in this work is about one order of magnitude larger than that obtained in Ref.~\cite{Ellis:2020ena}. 
	
	The anisotropies in the SGWB result from the large-scale perturbations of $\Phi$ which leave the Hubble horizon shortly after $T_{i}$. The averaged value of $\Phi$ in each large scale region, $\bar{\Phi}(\mathbf{x})$, affects the string length density and GW production. In the regions with larger $|\bar{\Phi}|$, it is difficult for quantum fluctuations to kick $\Phi$ across the potential barrier so that the length density of cosmic strings becomes smaller and GWs are less produced, leading to large anisotropies in the SGWB. 
	Because of the weak angular resolution of the GW detectors, we only focus on small multiple $\mathtt{l}$ of the spherical harmonic expansion of the SGWB, 
	corresponding to the inhomogeneities of GW energy density at very large scales. 
	\begin{figure*}[t]
		\includegraphics[width=2.3in]{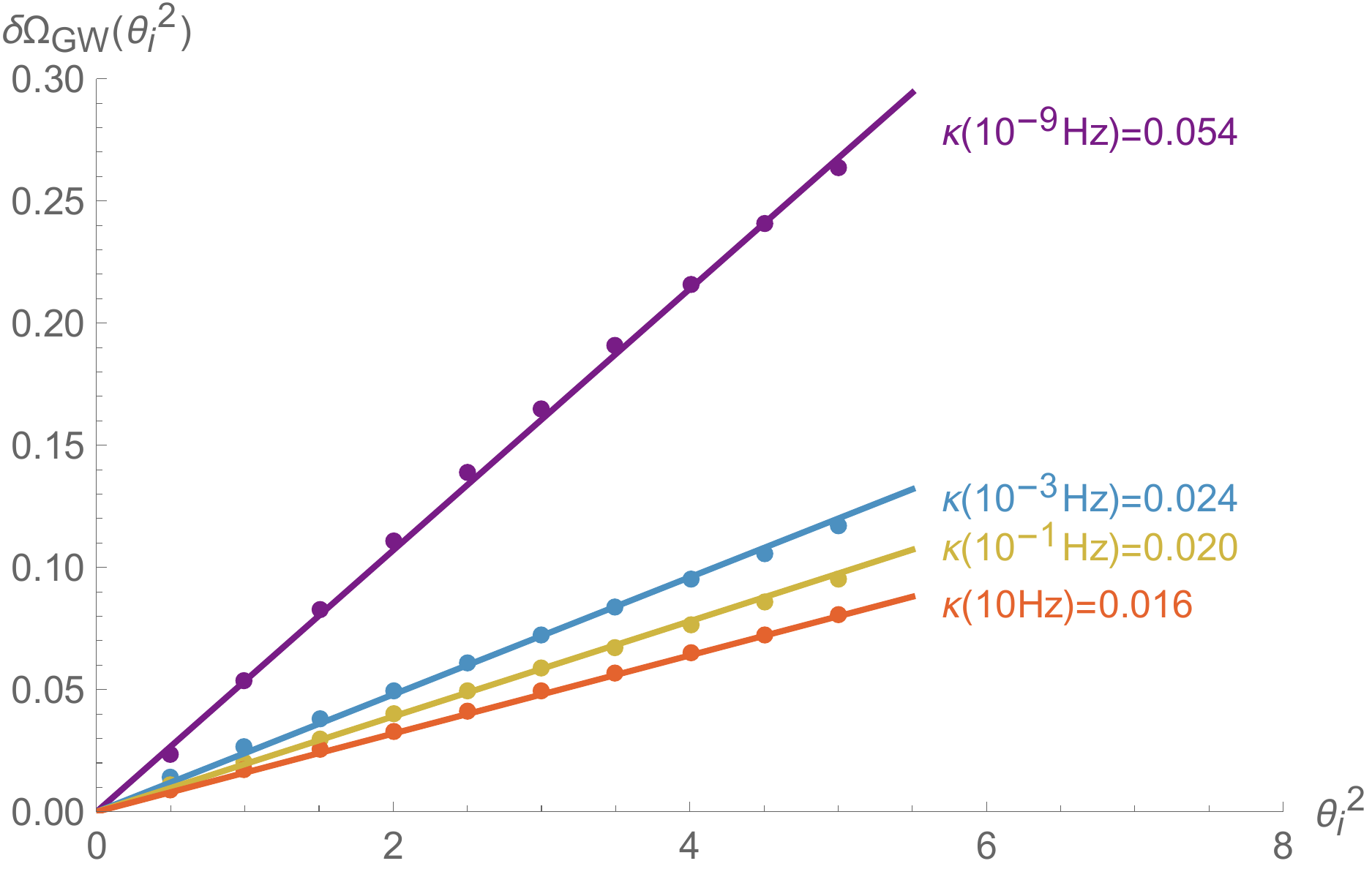}
		\includegraphics[width=2.3in]{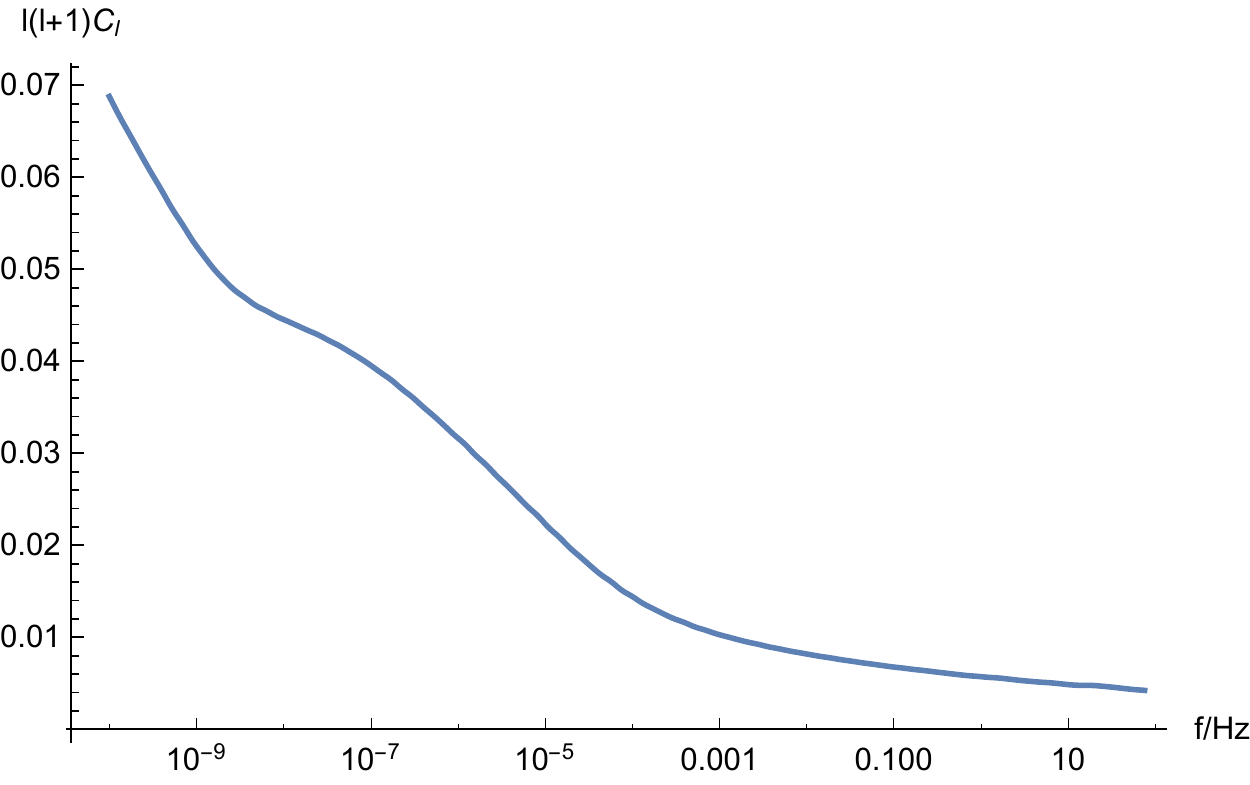}
		\includegraphics[width=2.3in]{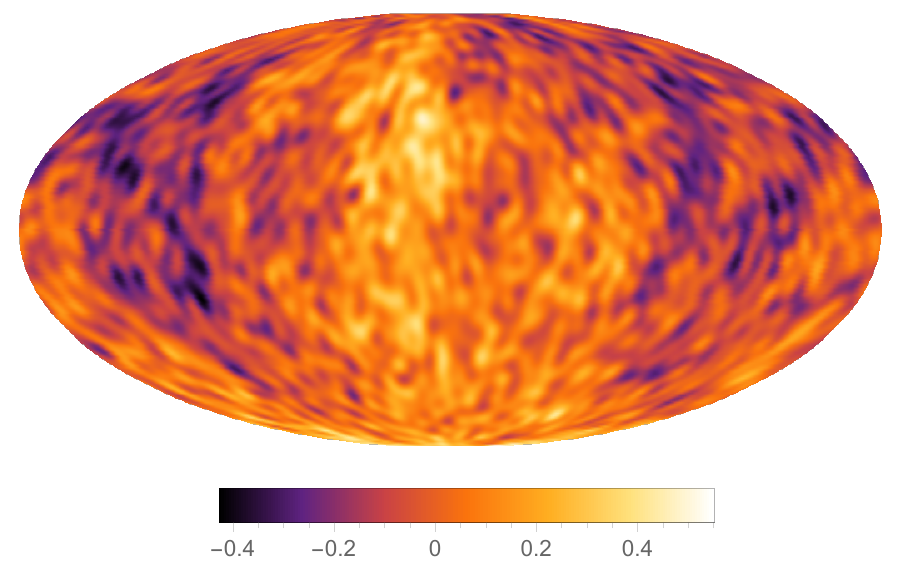}
		\caption{The left panel shows the numerical results of the linear dependence of $\delta\Omega_{\mathrm{GW}}$ on $\theta_{i}$ for $f=10^{-9}$Hz, $10^{-3}$Hz, $10^{-1}$Hz and $10$Hz. The middle panel shows the dependence of $\mathtt{l}(\mathtt{l}+1)C_{\mathtt{l}}$ on $f$. The right panel shows a stochastic realization of $\delta\Omega_{\mathrm{GW}}$ in the PTA sensitivity frequency band, where the angular power spectrum is $\mathtt{l}(\mathtt{l}+1)C_{\mathtt{l}}=0.056$ and we set the cutoff $\mathtt{l}_{\mathrm{cut}}=50$.
		}
		\label{fig:aniso}
	\end{figure*}
	Large scale perturbations of $\Omega_{\mathrm{GW}}$ is defined as
	\begin{equation} \label{eq:deltaOmega}
		\delta\Omega_{\mathrm{GW}}(f,  \mathbf{x})\equiv\frac{\Omega_{\mathrm{GW}}(f,\mathbf{x})-\Omega_{\mathrm{GW}}(f)}{\Omega_{\mathrm{GW}}(f)}\,.
	\end{equation}	
	Note that the GW energy spectrum is a function $\mathbf{x}$ in principle. Here we emphasize that $\Omega_{\mathrm{GW}}(f)$ in Eq.~\eqref{eq:Omega} is the averaged value over the whole Universe, while $\Omega_{\mathrm{GW}}(f,\mathbf{x})$ in Eq.~\eqref{eq:deltaOmega} is the averged value over a large scale region around $\mathbf{x}$.
	It is expected that $\delta\Omega_{\mathrm{GW}}(f,\mathbf{x})$ linearly depends on $|\bar{\Phi}(\mathbf{x})|^{2}$,
	\begin{equation} \label{eq:linear}
		\delta\Omega_{\mathrm{GW}}(f,\mathbf{x})=\kappa(f) \bar{\theta}^{2}(\mathbf{x})\,.
	\end{equation}	
	where $\bar{\theta}(\mathbf{x})\equiv\frac{|\bar{\Phi}(\mathbf{x})|}{H_{\mathrm{inf}}/2\pi}$ and $\kappa(f)$ is the frequency-dependent coefficient. Since we focus on  very large scales comparable to $H^{-1}_{0}$, $\kappa(f)$ can be numerically obtained from the dependence of $\Omega_{\mathrm{GW}}(f)$ on $\theta_{i}^{2}$. The left panel of Fig.~\ref{fig:aniso} verifies the linear assumption Eq.~\eqref{eq:linear} and $\kappa(f)$ is obtained numerically as $\kappa(10^{-9}\mathrm{Hz})=0.054$, $\kappa(10^{-3}\mathrm{Hz})=0.024$, $\kappa(10^{-1}\mathrm{Hz})=0.020$, $\kappa(10\mathrm{Hz})=0.016$. 
	
	The angular power spectrum at large scales can be expressed in terms of the power spectrum of $\delta \Omega_{\mathrm{GW}}(\mathbf{x})$ as $\mathtt{l}(\mathtt{l}+1)C_{\mathtt{l}}(f)=\frac{\pi}{2}\langle\delta\Omega_{\mathrm{GW}}^{2}(f,\mathbf{x})\rangle$~\cite{Liddle:2000cg}.
	Assuming perturbations of $\Phi$ is Gaussian,  Wick's theoriem implies that $\langle|\Phi^{2}(\mathbf{x})|^{2}\rangle=3\langle|\Phi^{2}(\mathbf{x})|\rangle$. Together with $\langle|\Phi^{2}(\mathbf{x})|\rangle=H_{\mathrm{inf}}^{2}/2\pi^{2}$, we finally obtain
	\begin{equation}\label{eq:angular}
	 \mathtt{l}(\mathtt{l}+1)C_{\mathtt{l}}(f)=6\pi\kappa^{2}(f)\,.  
	\end{equation}
	The middle panel of Fig.~\ref{fig:aniso} shows the dependence of $\mathtt{l}(\mathtt{l}+1)C_{\mathtt{l}}$ on frequency  $f$. In particular,  $\mathtt{l}(\mathtt{l}+1)C_{\mathtt{l}}(f)=5.6\times 10^{-2}$ for $f=10^{-9}$Hz~(PTA experiments), $1.1\times 10^{-2}$ for $f=10^{-3}$Hz~(LISA/Taiji), $7.6\times 10^{-3}$ for $f=10^{-1}$Hz~(DECIGO/BBO), $4.8\times 10^{-3}$ for $f=10$Hz~(LIGO/Virgo/KAGRA/CE/ET), respectively. 
	The right panel of Fig.~\ref{fig:aniso} shows a stochastic realization of anisotropies in the PTA sensitivity band, which is expected to be detected in the near future.
	

	
	\emph{Conclusion and discussion}. In this work, we find that large anisotropies are expected to present in the SGWB from cosmic strings associated with $U(1)$ symmetry breaking at the GUT scale. The string network is sparse and never reaches the scaling solution predicted by the VOS model. 
	The string loop distribution is determined by quantum fluctuations during inflation. 
	The SGWB from strings with tension $G\mu=4\times 10^{-10}$ can naturally explain the result with  the NANOGrav 12.5  data, and the anisotropies with augular power spectrum $\mathtt{l}(\mathtt{l}+1)C_{\mathtt{l}}=0.056$ are expected to be detected in SKA. 
	The frequency-dependent anisotropies can be tested by multiband observations of GWs.
	
	As the inflaton slowly rolls down the potential, the energy scale during inflation is not exactly a constant, which leads to a small deviation from our results $\mathtt{l}(\mathtt{l}+1)C_{\mathtt{l}}(f)$. This provides us a method to further distinguish inflationary models with more precise observations of the anisotropies in the SGWB. 
	
	
	In this work, we set the initial value $\Phi_{i}=0$ because the vacuum expectation value of $\Phi$ is zero before the $U(1)$ symmetry is broken. Quantum kicks before $t_{i}$ may cause a nonzero $\Phi_{i}$, especially for the models where inflation proceeds for a long time before $t_{i}$. In this case, the anisotropies become larger and carry the information of the inflationary scale. 
	

	
	\emph{Acknowledgments}
	We thank Ye-Ling Zhou for fruitful discussions.
	This work is supported in part by the National Key Research and Development Program of China Grant No. 2020YFC2201501, in part by the National Natural Science Foundation of China Grants
	No. 11690021, No. 11690022, No. 11851302, No. 11947302, No. 11991052, No. 11821505, No. 12105060 and No. 12105344, in part by the Science Research Grants from the China Manned Space Project with NO. CMS-CSST-2021-B01,
	in part by the Strategic Priority Research Program of the Chinese Academy of Sciences Grant No. XDB23030100, in part by the Key Research Program of the CAS Grant No. XDPB15 and by Key Research Program of Frontier Sciences, CAS.
	\bibliography{anisoCS}
\end{document}